\documentclass[10pt]{paper}

\begin{document}

\title{Chern-Simons Theory of Fractional Quantum Hall Effect in (Pseudo) Massless Dirac
      Electrons}
\author{Huabi Zeng \\
        Department of Physics, Nanjing University,\\
        Nanjing 210093, China\\
        zenghbi@gmail.com}

\maketitle

\begin{abstract}

We derive the effective field theory from the microscopic
Hamiltonian of interacting two-dimensional (pseudo) Dirac electrons
by performing a statistic gauge transformation. The quantized Hall
conductance are expected to be $\sigma_{xy}=\frac{e^2}{h}(2k-1)$
with $k$ is arbitrary integer. There are also topological
excitations which have fractional charge and obey fractional
statistics.

\end{abstract}

\section{Introduction}

The discovery of the fractional quantum hall effect\cite{1} in GaAs
have deepen our understanding of quantum many-body system very
much.\cite{2} The completely new matter state in FQHE is called
quantum hall liquid which opened a new chapter of condensed matter
physics. The quantum hall liquid has a new kind of order
(topological order) that is beyond the Landau symmetry-breaking
description.\cite{3} After laughlin's first successful theory in
which the famous laughlin wave function was proposed\cite{4}. A
completely and first principle construction of the
effective-field-theory was given by Zhang, Hansson and
Kivelson\cite{5,7} and later extended by Lee and Zhang\cite{6}. In
the building of the theory, a singular gauge transformation is used
to map the interacting fermions problem to one of interacting bosons
coupled to an additional gauge field (the Chern-Simons
field).\cite{7,8} This successful effective-field-theory not only
explains the experimental facts completely, but also demonstrates
that there is a deep connection between superfluid and FQHE. Until
now, the FQHE is only observed in GaAs, in which the electrons are
conventional (``non-relativistic``). A natural question is that what
is the properties of the possible FQHE in (pseudo) Massless Dirac
electrons that exist in Graphene.\cite{20} The possible FQHE in
Graphene has been discussed in many papers\cite{9,10,11,12,13}, and
until now, a effective field theory for FQHE in ``relativistic``
electrons system is still needed . In section 2, we try to derive
the effective-field-theory of FQHE from the full spin polarized
interacting microscopic massless Dirac fermions Hamiltonian by using
the same statistic gauge transformation in ref. 7. Then we consider
the uniform mean-field solution and the topological excitations of
the theory in section 3 and section 4 respectively. Finally,
conclusion is given in section 5.

\section{The Derivation of Chern-Simons Theory}
We begin by writing down the hamiltonian of interacting two
dimensional massless fermions \cite{14}
\begin{equation}
H=\sum_{i}v_{F}[\vec{\sigma}\cdot(\textbf{p}_{i}-\frac{e}{c}\textbf{A}(\textbf{r}_i))]+\sum_{i}eA_0(\textbf{r}_i)+\frac{1}{2}\sum_{i\neq
j}V(|\textbf{r}_i-\textbf{r}_j|).
 \end {equation}
$v_F$ is the fermi speed. We have assumed the (pseudo) Dirac
electrons are completely polarized. This Hamiltonian is different
from the one in Graphene, since there are two kinds of (pseudo) spin
while we only consider one kind here. The space wave function of the
Hamiltonian $\Psi(\textbf{r}_1,\textbf{r}_2....,\textbf{r}_N)$ must
be totally antisymmetric. The eigenvalue equation of the hamiltonian
reads
\begin{equation}
H\Psi(\textbf{r}_1,\textbf{r}_2....,\textbf{r}_N)=E\Psi(\textbf{r}_1,\textbf{r}_2....,\textbf{r}_N).
\end{equation}
$\Psi$ is a two components wave fuction
\begin{equation}
\Psi(\textbf{r}_1,\textbf{r}_2....,\textbf{r}_N)=[\begin{array}{c}
 \Psi_1(\textbf{r}_1,\textbf{r}_2....,\textbf{r}_N) \\
 \Psi_2(\textbf{r}_1,\textbf{r}_2....,\textbf{r}_N)
\end{array}].
\end{equation}
Both components are totally antisymmetric. This N-fermion problem
can be mapped to a bosonic eigenvalue one with symmetric wave
function by performing a statistic gauge transformation.\cite{5,7}
\begin{equation}
H'\Psi'(\textbf{r}_1,\textbf{r}_2....,\textbf{r}_N)=E\Psi'(\textbf{r}_1,\textbf{r}_2....,\textbf{r}_N),
\end{equation}
$\Psi'(\textbf{r}_1,\textbf{r}_2....,\textbf{r}_N)$ has two
components $\Psi_1', \Psi_2'$, both are completely symmetric when we
exchange the places of any two particles. In which
\begin{equation}
H'=U^{-1}HU, \Psi'=U^{-1}\Psi,
\end{equation}

\begin{equation}
U=Iexp(-i\sum_{i<j}\frac{\theta}{\pi}\alpha_{ij}).
\end{equation}
$I$ is the two dimensional identity matrix. $\alpha_{ij}$ is the
angel between the x-axis and the the vector
 $\textbf{r}_i-\textbf{r}_j$. $\theta=(2k+1)\pi$ must be satisfied to
guarantee $\Psi'$ is symmetric. By introducing the statistic gauge
operator
\begin{equation}
\textbf{a}(\textbf{r})=\frac{\phi_0\theta}{2\pi^2}\sum_{i\neq
j}\nabla\alpha_{ij},
\end{equation}
$\phi_0=\frac{2\pi}{e}$ is the flux quantum. By performing the
statistic gauge transformation (5). We can write the new bosonic
Hamiltonian
\begin{equation}
H'=\sum_{i}v_{F}[\vec{\sigma}\cdot(\textbf{p}_{i}-\frac{e}{c}\textbf{A}(\textbf{r}_i)-\frac{e}{c}\textbf{a}(\textbf{r}_i))]
+\sum_{i}eA_0(\textbf{r}_i)+\frac{1}{2}\sum_{i\neq
j}V(|\textbf{r}_i-\textbf{r}_j|).
\end {equation}
The physical meaning of the introduced statistic gauge field
$\textbf{a}$ can be seen from the statistic gauge transformation:
\begin{equation}
exp(i\sum_{i<j}\frac{\theta}{\pi}\alpha_{ij})\textbf{p}_iexp(-i\sum_{i<j}\frac{\theta}{\pi}\alpha_{ij})=\textbf{p}_i-\frac{e}{c}\frac{\phi_0\theta}{2\pi^2}\sum_{i\neq
j}\nabla\alpha_{ij}=\textbf{p}_i-\frac{e}{c}\textbf{a},
\end{equation}
$\textbf{a}$ describes the gauge interaction between particles.

In the language of second quantization, the Hamiltonian can be
rewritten as
\begin{equation}
H'=\int
d^2\textbf{r}\phi^+(\textbf{r})[eA_0+v_F(\vec{\sigma}\cdot(\textbf{p}_{i}-\frac{e}{c}\textbf{A}(\textbf{r}_i)-\frac{e}{c}\textbf{a}(\textbf{r}_i))]\phi(\textbf{r})+\frac{1}{2}\int
d^2\textbf{r}\int
d^2\textbf{r}'(\rho(\textbf{r})-\bar{\rho})V(\textbf{r}-\textbf{r}')(\rho(\textbf{r}')-\bar{\rho}),
\end{equation}
where $\rho(\textbf{r})=\phi^+(\textbf{r})\phi(\textbf{r})$ is the
particle density at $\textbf{r}$. $\bar{\rho}$ is the average
particle density.
The statistic operator expressed in second
quantization reads
\begin{equation}
a^{\alpha}=\frac{\phi_0\theta}{2\pi^2}\varepsilon^{\alpha\beta}\int
d^2\textbf{r}'\frac{\textbf{r}^\beta-\textbf{r}'^\beta}{|\textbf{r}-\textbf{r}'|^2}\rho(\textbf{r}').
\end{equation}
$\varepsilon^{\alpha\beta}=\varepsilon^{0\alpha\beta}$ is the
standard Levi-Civita tensor. From (11) we know that $\textbf{a}$ is
decided by $\rho(\textbf{r})=\phi^+(\textbf{r})\phi(\textbf{r})$, it
is not an independent dynamic quantity. We must know the equation of
motion for $\textbf{a}$ in our theory to get the effective action of
the system. In the Coulomb gauge, (11) is the solution of
\begin{equation}
\varepsilon^{\alpha\beta}\partial_{\alpha}a_{\beta}(\textbf{r})=\phi_0\frac{\theta}{\pi}{\rho}(\textbf{r}).
\end{equation}
Equation just give $\textbf{a}(\textbf{r})$ at a given time. In
order to get the dynamics of the statistic gauge field, we take the
time derivation of (12)
\begin{equation}
\varepsilon^{\alpha\beta}\partial_{\alpha}\dot{a}_{\beta}(\textbf{r})=\phi_0\frac{\theta}{\pi}\dot{\rho}(\textbf{r}).
\end{equation}
By using the continuity equation
$\partial_t\rho(\textbf{r},t)+\partial_{\alpha}j^{\alpha}(\textbf{r},t)=0$,
we have
\begin{equation}
\varepsilon^{\alpha\beta}\dot{a}_{\beta}(\textbf{r})=-\phi_0\frac{\theta}{\pi}j^{\alpha}.
\end{equation}
Equation (12) and (14) together give the equations of motion for
$\textbf{a}$, they can be derived from the Chern-Simons lagrange
\begin{equation}
\mathcal{L}=\frac{1}{2}\frac{e\pi}{\phi_0\theta}\varepsilon^{\mu\nu\rho}a_\mu\partial_\nu
a_\rho-a_{\mu} j^{\mu} .
\end{equation}

Then we can formulate the problem in coherent state path integral,
all the thermodynamic properties and the electromagnetic response of
the system is completely contained in the following partition
function $Z$
\begin{equation}
Z[A_\mu]=\int[da_\mu][d\phi]exp(iS[a_\mu]+iS[\phi]).
\end{equation}
In which
\begin{equation}
S[a_\mu]=\int dt \int d^2\textbf{r}\frac{\pi e}{2\theta
\phi_0}\varepsilon^{\mu\nu\rho}a_\mu\partial_\nu a_\rho,\\
\end{equation}

\begin{equation}
 S[\phi]=\int dt\int d^2\textbf{r}\phi^+[\textbf{r})(i\partial_t-\frac{e}{c}(A_0+a_0)-\vec{\sigma}\cdot(\textbf{p}-\frac{e}{c}(\textbf{A}+\textbf{a}))]\phi(\textbf{r})-\int dt \int d^2\textbf{r} \int
 d^2\textbf{r}'\delta\rho(\textbf{r}')V(\textbf{r}-\textbf{r}')\delta\rho(\textbf{r}).
\end{equation}
$\phi$ also has two components. It is clear that the action has a
local U(1) gauge symmetry

\begin{equation}
\phi\rightarrow \phi'= (Iexp(i\gamma))\phi,  a_\mu\rightarrow
a_\mu'=a_\mu-\frac{1}{e}\partial_\mu \gamma.
\end{equation}
$I$ is the two dimensional unit matrix.

\section{The Mean Field Theory}
First we consider the mean-field solution of the system when there
is no external electric field ($A_0=0$). For the perpendicular
external magnetic field along the $z$ axis,
$\varepsilon^{\alpha\beta}\partial_{\alpha}A_{\beta}=-B$. We can
guess the mean-field solution
\begin{equation}
\phi(\textbf{r})=(\begin{array}{c}
 \sqrt{\bar{\rho}}_1 \\
 \sqrt{\bar{\rho}}_2
\end{array}), a(\textbf{r})=-A(\textbf{r}),a_0(\textbf{r})=0.
\end{equation}
 Both $\bar{\rho_1}$ and$\bar{\rho_2}$ are constant, they are not independent of each other since $\bar{\rho}_1+\bar{\rho}_2=\bar{\rho}$
  is the average electron density. It is easy to prove that
the solution satisfy all the equations of motion derived from the
action. Further more, the statistic gauge field is related to the
particle density via (12), we get
\begin{equation}
B=\phi_0\frac{\theta}{\pi}\bar{\rho},
\end{equation}
 which means that the filling factor
\begin{equation}
 \nu=\frac{\pi}{\theta}=\frac{1}{2k-1}.
\end{equation}
In this special state, we can consider a bosonic system without a
magnetic field where a Bose condensation or superfluid will happen.

Then we calculate the Hall conductance, we apply an external scalar
potential $A_0$ with $\partial_\mu A_0=-E_\mu$. The gauge invariant
current
\begin{equation}
\langle j_\alpha(\textbf{r}) \rangle=\langle \frac{\delta S}{\delta
A_\alpha}\rangle.
\end{equation}
From the action, we have
\begin{equation}
  j_\alpha(\textbf{r})=\frac{\delta S}{\delta A_\alpha}=\frac{\delta S_\phi}{\delta
 a_\alpha}=-\frac{\delta S_a}{\delta
 a_\alpha}.
\end{equation}
In the last step of (24) we have used the static field equation that
$\frac{\delta S}{\delta a_\alpha}=0$. After integration by parts the
Chern-Simons lagrangian can be written as
\begin{equation}
\mathcal{L}_a=\frac{e\pi}{2\theta\phi_0}\epsilon^{\alpha\beta}(2a_\alpha\partial_\beta
a_0-a_\alpha\partial_t a_\beta).
\end{equation}
Since the statistic gauge field $\textbf{a}$ is static, we finally
get
\begin{equation}
j_\alpha=\frac{e^2\pi}{h\theta}\epsilon^{\alpha\beta}E_\beta.
\end{equation}
 The expectation of the current
\begin{equation}
\langle j_\alpha(\textbf{r})
\rangle=\frac{1}{Z}\int[d\phi][da_\alpha](\frac{e^2\pi}{h\theta}\epsilon^{\alpha\beta}E_\beta)e^{iS[a_\mu]+iS[\phi]}.
\end{equation}
In the mean field theory, we can replace $S[\phi]+S[a_\mu]$ by $S$
which computed by using the classic path. Then the fields equate to
their expectation value. The current equals to
\begin{equation}
\langle j_\alpha
\rangle=j_\alpha=\frac{e^2\pi}{h\theta}\varepsilon^{\alpha
\beta}E_\beta.
\end{equation}
Which means that the Hall conductance
\begin{equation}
\sigma_{xx}=0, \sigma_{xy}=\frac{e^2}{h}\frac{1}{2k-1}.
\end{equation}
Where $k$ is an arbitrary integer, the Hall conductance is
quantized. And the odd fraction should be the same as the
``non-relativistic`` electrons. Due to the Anderson-Higgs mechanism,
the non-zero vacuum break the gauge symmetry results in the Meissner
effect. The effect leads the state to be a incompressible quantum
fluid, since any change of electron density will change the
statistic gauge field which results in some flux that is forbidden
by the Meissner effect.

\section{The Vortices}
Similar to the conventional electrons system, in addition to the
uniform ground state there exist static, non-uniform, finite energy
vortex solutions (topological excitations). The asymptotically
behavior at ($\textbf{r} \rightarrow \infty$) of the solution is
\begin{equation}
\phi(\textbf{r})=(\begin{array}{c}
 \sqrt{\bar{\rho}}_1 \\
 \sqrt{\bar{\rho}}_2
\end{array})e^{\pm i\psi(\textbf{r})},
\end{equation}

\begin{equation}
\delta a=\mathbf{a}+\mathbf{A}=\pm\frac{1}{e}\nabla
\psi(\mathbf{r}).
\end{equation}
Where $\psi$ is the angle of $\textbf{r}$. Therefore for a large
contour we have
\begin{equation}
\oint \delta a\cdot d\textrm{l}=\pm2\pi/e=\pm\phi_0
\end{equation}
Rather the quantization of flux, the charge is fractional quantized.
To see this, form (12),
\begin{equation}
\rho=\bar{\rho}+\delta
\rho=\frac{\nu}{\phi_0}\epsilon^{\alpha\beta}\partial_{\alpha}a_{\beta}=\frac{\nu}{\phi_0}\epsilon^{\alpha\beta}\partial_{\alpha}(\delta
a-A_{\beta})=\frac{\nu}{\phi_0}\epsilon^{\alpha\beta}\partial_{\alpha}\delta
a_{\beta}+\frac{\nu}{\phi_0}B.
\end{equation}
Therefor the excess charge of the vortex
\begin{equation}
Q=e\int d^2\textbf{r}\delta
\rho(\textbf{r})=e\frac{\nu}{\phi_0}\oint \delta a\cdot
d\textrm{l}=\pm e\nu.
\end{equation}
This demonstrates that at the fractional filling the topological
excitation have fractional charge. These field configurations
corresponding to the quasiparticles and quasiholes above the ground
state with fractional charge. The fractional charge implies that the
qusiparticles and qusiholes are anyons obey fractional statistics
with $\theta_1=\pi\nu$.\cite{15,16,17,18,19}

\section{Conclusion}
In this paper, we find that the statistic gauge transformation can
also be applied to the interacting (pseudo) Dirac electrons system.
In the Gindzburg-Landau-Chern-Simons theory proposed in ref. 5, the
field coupled to the Maxwell field and the statistic gauge field is
a bosonic complex scalar field, while in the (pseudo) Dirac
electrons system, the bosonic field becomes a two dimensional
complex scalar field. By considering the mean field solution of the
effective field theory. We find similar results to
``non-relativistic`` electrons: the fractional Hall conductance with
odd denominators, the fractional charge of the quasiparticles. So we
conclude that although the linear dispersion of (pseudo) Dirac
electrons is different from that of the conventional electrons, the
phenomenon of FQHE are also expected to be similar to the
conventional electrons.

\section{Acknowledgement}
I thank professor A.zee for his great encouragement. I also thank
professor Hongshi Zong for his great help.

\end{document}